\documentclass[useAMS,usenatbib]{mnras}

\usepackage{url,times,hyperref,graphicx,amsmath,amsfonts,amssymb,aas_macros,color,epsfig,epstopdf}

\newcommand{\mhalo}{M$_{\rm halo}$}
\newcommand{\mstar}{M$^{\star}$}
\newcommand{\lcdm}{$\Lambda$CDM}
\newcommand{\msun}{M$_\odot$}

\newcommand{\hMpc}{{\ifmmode{h^{-1}{\rm Mpc}}\else{$h^{-1}$Mpc}\fi}}
\newcommand{\hkpc}{{\ifmmode{h^{-1}{\rm kpc}}\else{$h^{-1}$kpc}\fi}}
\newcommand{\hMsun}{{\ifmmode{h^{-1}{\rm {M_{\odot}}}}\else{$h^{-1}{\rm{M_{\odot}}}$}\fi}}
\newcommand{\ltsima}{$\; \buildrel < \over \sim \;$}
\newcommand{\gtsima}{$\; \buildrel > \over \sim \;$}
\newcommand{\lsim}{\lower.5ex\hbox{\ltsima}}
\newcommand{\gsim}{\lower.5ex\hbox{\gtsima}}
\def\lcdm{$\Lambda$CDM}

\def\lesssim{\mathrel{\hbox{\rlap{\hbox{\lower4pt\hbox{$\sim$}}}\hbox{$<$}}}}
\def\gtrsim{\mathrel{\hbox{\rlap{\hbox{\lower4pt\hbox{$\sim$}}}\hbox{$>$}}}}

\newcommand{\Fig}[1]{Fig.~\ref{#1}}
\newcommand{\beq}{\begin{equation}}
\newcommand{\eeq}{\end{equation}}
\def\beqa{\begin{eqnarray}}
\def\eeqa{\end{eqnarray}}
\def\hMpc{$h^{-1}\,{\rm Mpc}$}
\def\hkpc{$h^{-1}\,{\rm kpc}$}
\def\lcdm{\ensuremath{\Lambda}CDM}

\def\head{
 \vbox to 0pt{\vss
                   \hbox to 0pt{\hskip 440pt\rm LA-UR-10-07069\hss}
                  \vskip 25pt}}

\title[Mass discrepancy validates \lcdm]
{The mass discrepancy acceleration relation in a \lcdm\ context}
\author[Di Cintio]
       {Arianna Di Cintio$^{1}$\thanks{E-mail: arianna.dicintio@dark-cosmology.dk} \& Federico Lelli$^{2}$\\
$^{1}$Dark Cosmology Centre, Niels Bohr Institute, University of Copenhagen, Juliane Maries Vej 30, DK-2100 Copenhagen, Denmark\\
$^{2}$Astronomy Department, Case Western Reserve University, 10900 Euclid Avenue, Cleveland, OH 44106, USA\\}

\begin{document}

\date{Accepted 2015 November 19. Received 2015 November 19; in original form 2015 October 23}

\pagerange{\pageref{firstpage}--\pageref{lastpage}} \pubyear{2010}

\maketitle

\label{firstpage}


\begin{abstract}
The mass discrepancy acceleration relation (MDAR) describes the coupling between baryons and dark matter (DM) in galaxies: the ratio of total-to-baryonic mass  at a given radius anti-correlates with the acceleration due to baryons. The MDAR has been seen as a challenge to the \lcdm\ galaxy formation model, while it can be explained by Modified Newtonian Dynamics. In this \textit{Letter} we show that the MDAR arises in a \lcdm\ cosmology once observed galaxy scaling relations are taken into account. We build semi-empirical models based on \lcdm\ haloes, with and without the inclusion of baryonic effects, coupled to empirically motivated structural relations. Our models can reproduce the MDAR: specifically, a mass-dependent density profile for DM haloes can fully account for the observed MDAR shape, while a universal profile shows a discrepancy with the MDAR of dwarf galaxies with \mstar$<$$10^{9.5}$\msun, a further indication suggesting the existence of DM cores. Additionally, we reproduce slope and normalization of the baryonic Tully-Fisher relation (BTFR) with 0.17 dex scatter. 

These results imply that in \lcdm\ (i) the MDAR is driven by structural scaling relations of galaxies and  DM density profile shapes, and (ii) the baryonic fractions determined by the BTFR are consistent with those inferred from abundance-matching studies.

\end{abstract}

\noindent
\begin{keywords}
 galaxies: evolution - formation - haloes cosmology: theory - dark matter
 \end{keywords}

\section{Introduction} \label{sec:introduction}
Disc galaxies show a tight coupling between their observed dynamics and the distribution of baryons \citep[e.g.][]{Sancisi04,swaters12,Lelli13}. This is quantified by the mass discrepancy acceleration relation (MDAR). Following \citet{mcgaugh04,mcgaugh14} the mass discrepancy is defined as V$\rm_{obs}^2(r)$/V$\rm_{bar}^2(r)$, where V$\rm_{obs}^2$ is the total observed rotational velocity and V$\rm_{bar}^2$ is the  rotation velocity inferred from the distribution of baryons (gas and stars). For a spherical mass distribution, V$\rm_{obs}^2(r)$/V$\rm_{bar}^2(r)$=$\rm M_{tot}(r)/M_{bar}(r)$.

\citet{mcgaugh04,mcgaugh14} finds that a very tight correlation exists between the mass discrepancy and the gravitational acceleration due to  baryons $\rm g_{bar}$=$\rm V_{bar}^2(r)/r$. The existence of the MDAR has been  difficult to reconcile with a standard \lcdm\ picture, hinting perhaps toward a fundamental law of nature \citep{mcgaugh14} and finding a natural explanation in Modified Newtonian Dynamics (MOND) theory \citep{milgrom83}, which posits a characteristic acceleration scale.
Some authors have previously pointed out that a correlation between mass discrepancy and baryonic acceleration can be expected in \lcdm\ once dissipative processes of galaxy formation are properly accounted for.
\citet{vdb00} used a semi-analytic model for the formation of galaxies, tuned to fit the Tully-Fisher relation, to show that  a characteristic acceleration can be reproduced in \lcdm\ models.
  Yet a more challenging exercise is to verify whether simulations and models including baryon physics can reproduce the small scatter about the observed MDAR  \citep{walker14}. Recently, \citet{santos15} have shown that a suite of simulated galaxies  from the MaGICC project \citep{brook12,stinson13} can successfully reproduce both the MDAR and its scatter. Those simulations also reproduce observed galaxy scaling relations and  give rise to the expected baryonic Tully-Fisher relation (BTFR), i.e. the relation between baryonic mass and galaxy velocity \citep{mcgaugh12BAR}.

In this work, we explore the issue using our current knowledge of observations and theory: we derive the MDAR and the BTFR using a semi-empirical-\lcdm-based model and we compare them against a dataset of 146 galaxies (Lelli et al. in prep).
We build a  population of model galaxies whose distribution in stellar mass and other structural properties follows closely the observational sample, allowing us to explore the MDAR and BTFR. To this aim we use empirical and theoretical scaling relations superimposed on dark matter (DM) haloes whose density profiles are predicted in \lcdm\  with and without baryonic effects.

For the DM distribution we employ the universal NFW \citep{navarro96} profile and the mass-dependent density profile DC14 \citep{dicintio2014a,dicintio2014b}. The latter takes into account the core formation process resulting from impulsive outflows of gas in star forming region, which has been observed in hydrodynamical simulations at z=0 and across cosmic time \citep{governato10,teyssier13,dicintio2014a,onorbe15,chan15,TG15,tollet15}, and confirmed by analytic models \citep{pontzen12}.

The DC14 profile has a peak in \textit{core formation efficiency} for galaxies of \mstar$\sim10^{8.5}$\msun, a result largely confirmed by the increased sample of galaxies in  \citet{tollet15} as well as by \cite{chan15}. Such profile has been applied to a variety of studies, having proven its validity in the context of  velocity function of galaxies, Local Group predictions and rotation curves shapes \citep{brook15a,brook15b,brook15c}.
Once a DM halo is set, stellar and gaseous components are added using observed galaxy scaling relations. Theoretical rotation curves and mass profiles are then derived, allowing us to evaluate both the MDAR and the BTFR and to compare them with observations.

\section{Methods} \label{sec:model}
  \begin{figure}
\hspace{-.1in}\includegraphics[width=3.5in,height=3.in]{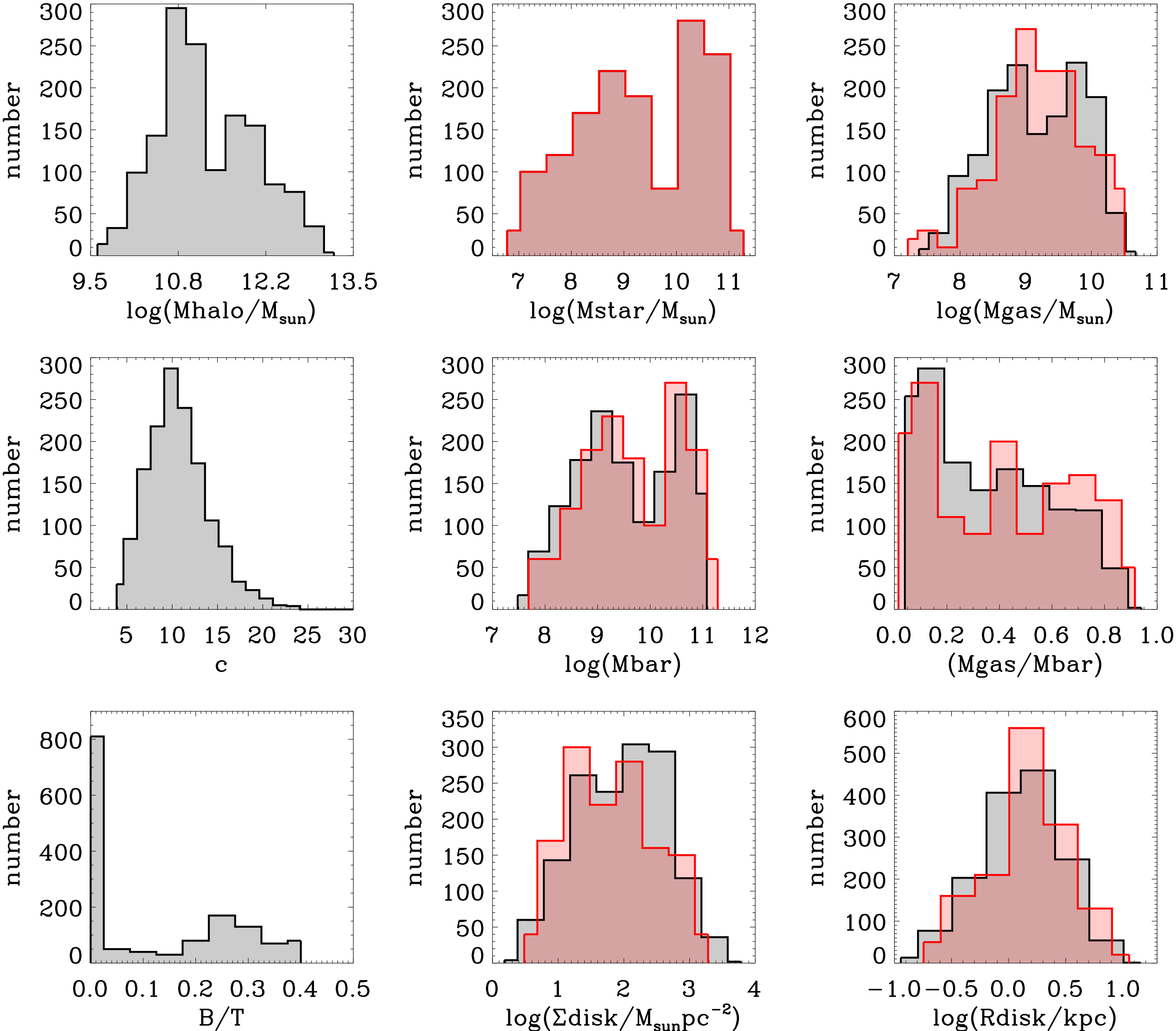}\caption{Properties of the observational (red) and theoretical  (black) sample of galaxies. From left to right  we show the distribution in halo, stellar and gas mass, halo concentration, baryonic mass, gas fraction, bulge-to-total ratio, central surface brightness and stellar disk scale length.}
\label{fig:sample} 
\end{figure}

   \begin{figure}
\vspace{-0.45cm} \includegraphics[width=3.3in,height=2.3in]{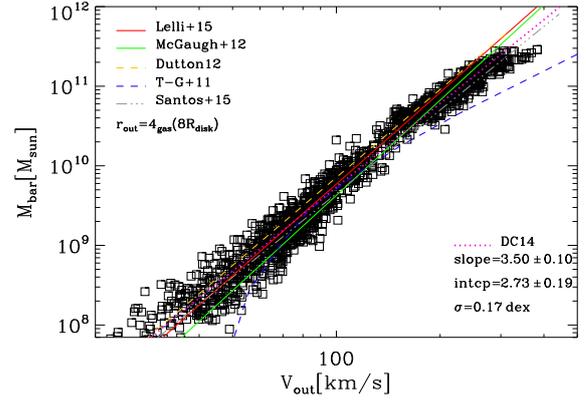} \caption{Baryonic Tully-Fisher relation derived by measuring the rotation of  model galaxies at $\rm 8R_{disk}$=$\rm 4R_{gas}$  (black squares). The best fit is indicated as dotted pink line for the DC14 profile. BTFRs from the literature are shown as solid, dashed and dotted-dashed line for observations, \lcdm-based models and hydro simulations, respectively.}
\label{fig:btf} 
\end{figure}
 
 \subsection{Observational sample}
The observed MDAR  is derived from the SPARC (Spitzer Photometry \& Accurate Rotation Curve) dataset (Lelli et al. in prep.). For this study, we consider 146 galaxies with high-quality HI/H$\alpha$ rotation curves and inclination angles above 30 degrees, in order to exclude face-on galaxies that require large inclination corrections on V$\rm_{obs}$. The mass models will be presented in detail in Lelli et al. (in prep.). In short, the baryonic contribution to the rotation curve is calculated from the observed surface density profiles of stars and gas by solving the Poisson equation in cylindrical coordinates \citep{casertano83}. 
The amplitude of $\rm V_{star}$ is scaled to the total stellar mass assuming a constant mass-to-light ratio of 0.5 at 3.6 $\mu$m for all galaxies \citep{mcgaughschombert14}. Similarly, the amplitude of $\rm V_{star}$ is scaled to the total gas mass as $\rm \sqrt{1.33} M_{HI}$, where $\rm M_{HI}$ is the observed HI mass and the factor 1.33 takes the contribution of Helium into account.
The galaxies cover a broad range in baryonic masses, from gas-dominated dwarfs with \mstar$\sim10^7$\msun\ to star-dominated spirals with \mstar$\sim10^{11}$\msun. It also spans a broad range in surface brightnesses, including many low-surface-brightness galaxies that are often misrepresented in optical surveys \citep[e.g.][]{mcgaugh95}. The statistical properties of the observational sample are shown in \Fig{fig:sample} as red histograms.

 \subsection{Theoretical model}

\noindent We build $\sim$1500 model galaxies such that their distribution in stellar mass follows the observed one, as shown in the top-middle panel of \Fig{fig:sample}. Each stellar mass bin is populated with 10$\times$  the observed number of galaxies, in order to have better statistic. The observational histograms (red) are rescaled to be compared with our theoretical sample.
Each galaxy is then assigned to a  DM halo using the abundance matching relation of \citet{kravtsov14} with a scatter of 0.10 dex.
Such abundance matching is formally identical to the \citet{moster13} one at the low mass end, but different for \mhalo$>$$10^{11.5}$\msun, such that a galaxy of \mstar$\sim$$2$$\times$$10^{11}$\msun\ is assigned to a $10^{13}$\msun\ halo rather than to a $10^{14}$\msun\ one. 
 The steeper slope of this relation at  high masses, based on stellar mass function calibrations  from  improved photometry \citep{bernardi13}, shows an excellent agreement  with recent hydrodynamical simulations \citep{tollet15}. 
We find that the less massive haloes implied by the \citet{kravtsov14} relation better reproduce the rotation curves of  galaxies with \mstar$\gtrsim 5$$\times$$10^{10}$\msun\ than the \citet{moster13} relation does. The resulting distribution in halo masses is shown in the top left panel of \Fig{fig:sample} and covers the range $\rm 9.60$$<$$\rm{log}_{10}$$($\mhalo/\msun$)$$<$$13.35$.

Our model galaxies live in DM haloes that follow the concentration-mass relation  from N-body simulations, assuming Planck cosmology \citep{dutton14}:\beq
\rm log_{10}(c_{200})=0.905-0.101log_{10}(M_{200}/[10^{12}h^{-1}M_{\odot}])
\eeq
\noindent  where c$_{\rm 200}$= R$_{\rm 200}$/r$_{\rm s}$, with a scale radius r$_{\rm s}$, and the mass is defined as M$_{\rm 200}$=M$_{\rm halo}$=R$_{\rm 200}\rm\Delta H^2$/2G, with $\Delta$$\sim$$200$. The c-M$_{\rm halo}$ relation has an intrinsic scatter of 0.11 dex.

DM haloes are described by two different density profiles: the well know universal, centrally-cuspy NFW one \citep{navarro96}, as found in collisionless simulations, and the DC14 model \citep{dicintio2014a,dicintio2014b}, derived from hydro simulations of galaxy formation that take into account a realistic treatment of baryonic physics \citep{stinson13,brook12}.
The DC14 profile is a double-power law one, whose main parameters (namely inner, $\gamma$, and outer, $\beta$, slope and sharpness of the transition, $\alpha$) depend on the integrated stellar-to-halo mass ratio of each galaxy, according to Eq. 3 of \citet{dicintio2014a}: it is able to describe at the same time cuspy ($\gamma$$\geqslant$1) as well as cored ($\gamma$$<$1) profiles, providing a useful parametrization for DM  haloes whose inner structure have ben modified by baryonic processes. Once ($\alpha,\beta,\gamma$) are specified via the quantity \mstar/\mhalo, the DC14 profile has the same number of free parameters as the NFW model, i.e. the total halo mass \mhalo\ and the scale radius $\rm r_s$.

Dwarf galaxies with M$^{\star}$/M$_{\rm halo}$$<$$10^{-4}$ do not have sufficient energy from their stellar component to generate powerful gas outflows that  cause the central DM to expand, hence they retain the initial NFW profile. More massive galaxies in the range M$^{\star}$/M$_{\rm halo}$$\sim$$3$$-$$5$$\times$$10^{-3}$ will instead have a central distribution of DM shallower than NFW: the core formation mechanism  is  most efficient at such mass range. Galaxies with \mstar/\mhalo$\geqslant$0.05, like the Milky Way (MW), will return to have a cuspy NFW profile due to the inefficient effect of SN feedback in contrasting the deep gravitational potential. At the MW mass, a contraction is applied as found in cosmological simulations \citep{dicintio2014a,tollet15}.
Since the DC14  profile has only been tested up to \mhalo$=$$10^{12}$\msun, we assume that more massive haloes retain a NFW profile with usual concentration.

For each model galaxy, the gas mass is assigned using the empirical M$\rm_{HI}$-M$^{\star}$ scaling relation obtained  by \citet{papastergis12}, which made use of the SDSS \citep{aba09} and the ALFALFA \citep{Haynes11} survey to determine the atomic hydrogen vs stellar mass of galaxies with a scatter of $\sim$0.20 dex.:
\beq
 \rm log_{10}(M_{\rm HI}/M^{\star})=-0.43log_{10}(M^{\star})+3.75
 \eeq
 \noindent  The total gas mass is given by M$_{\rm gas}$=1.33M$_{\rm HI}$, neglecting the minor contribution of molecular gas as in the SPARC mass models.
 
Stars and gas are distributed in a disk with an exponential profile. The stellar disk scale length, R$_{\rm disk}$, is assigned using the mass-size relation obtained by the GAMA survey in the K-band for late-type galaxies at z$<$$0.1$ \citep{Lange15}:
\beq
\rm R_{eff}=0.13(M^{\star})^{0.14}(1.0+\frac{M^{\star}}{14.03\times10^{10}M_{\odot}})^{0.77}
\eeq

\noindent where R$_{\rm eff}$=1.678R$_{\rm disk}$ for en exponential disk.
This equation is derived considering galaxies with \mstar$>$$10^9$\msun\ and its extrapolation to lower masses does not correctly reproduce the properties of our sample. Hence, for \mstar$<$$10^9$\msun, we use the best-fit relation to the dwarf galaxies in our sample: $\rm log_{10}(R_{eff})$$=$$\rm (log_{10}(M^{\star})$$-$$7.79)/2.47$.
The scatter in both relations is $\sim$0.20dex. The gas disk scale length is  R$_{\rm gas}$=2R$_{\rm disk}$.

Galaxies with \mstar$>$$10^{9.5}$\msun\ have a central bulge which follows a spherically symmetric  \citet{hernquist90} profile:
\beq
\rm \rho(r)=\frac{M_{bulge}R_{1/4}}{2\pi r(r+R_{1/4})^3}
\eeq

\noindent where R$_{1/4}$ is the radius at which the bulge mass is equal to M$_{\rm bulge}/4$. The half-mass radius, $\rm R_{1/2}$$=$$(1+\sqrt{2})R_{1/4}$, is gauged from the $\rm R_{1/2}$$-$$\rm M_{bulge}$ scaling relation as in \citet{Gadotti09}:
\beq
\rm log_{10}(R_{1/2})=0.30log_{10}(M_{bulge})-3.124
\eeq

\noindent We impose a linear bulge growth with stellar mass, $\rm M_{bulge}/M^{\star}$$=$$\rm (log_{10}(M^{\star})$$-$$9.5)/4.2$, such that the bulge-to-total ratio (B/T) is  $\sim30\%$ at the  MW mass, and further the relations between B/T vs \mstar, $\rm R_{1/2}$ and Hubble Type match this and other observational samples \citep{Gadotti09,moran12}.

\noindent The previous relations use a \citet{chabrier03} initial mass function and units of \msun\ and kpc. Further, we assume that their scatters are uncorrelated. In \Fig{fig:sample} the black histograms show that such semi-empirical model can reproduce very well the statistical properties of the SPARC observational sample of galaxies.
  
The total circular velocity of a halo following either a NFW or a DC14 density profile, embedding a galaxy with an exponential disk  of gas and stars and a central stellar bulge, reads:
\begin{equation}
 \rm{V}_{tot}^2(r)={ \rm{V}_{disk,g}^2(r) + \rm{V}_{disk,\star}^2(r) + \rm{V}_{bulge}^2(r) +  \rm{V}_{DM}^2(r)}
\end{equation}

\noindent where  V$_{\rm DM}$ has been multiplied by a factor $\sqrt{1-f_b}=\sqrt{0.842}$ to take into account the cosmic baryon fraction.
 The potential of a thin disk of mass M$_{\rm disk}$ and scale length R$_{\rm disk}$ is  \citep{Freeman70}:
\begin{equation}
\rm{V}_{disk}^2(r)=\frac{GM_{disk}}{R_{disk}}2y^2[I_0(y)K_0(y) - I_1(y)K_1(y)]
\end{equation}
\noindent where $\rm I_n$ and $\rm K_n$ are modified Bessel functions of the first and second kind, respectively, and  y=r/(2R$_{\rm disk}$). The potential of the bulge and DM halo are computed assuming spherical symmetry.
The velocity and the mass profile M(r) of each galaxy component can then be evaluated at every radius.
We remark here that our model generates realistic rotation curves of galaxies at different masses, a result that will be presented in a companion paper.

\section{Results} \label{sec:results}
We start by showing that our model is able to reproduce the BTFR, which is the relation between total baryonic mass versus galaxy rotation velocity along the flat part of the rotation curve (V$\rm_{flat}$) \citep{mcgaugh05,mcgaugh12BAR}. We measure the model rotation curves at $\rm 8R_{disk}$, such that we are encompassing the radius within which most of the baryonic mass is found. We note that the radius at which model rotation curves are measured is critical in the definition of the BTFR, resulting in different scatter and slopes at the low mass end, issue that will be explored in a forthcoming paper. 

The model galaxies are shown in \Fig{fig:btf} as black squares for a DC14 profile, with the best linear fit indicated as dotted pink line. The resulting slope, intercept and scatter about the average relation, for equally weighted points, are 3.50$\pm$0.10, 2.73$\pm$0.19 and 0.17 dex. Measuring at $\rm 8R_{disk}$ produces only a small difference in the BTFR between a DC14 and NFW model (thus not shown here), since this radius is large enough to have reached the point at which the two profiles are similar. The slope, intercept and scatter in the NFW case would read 3.74$\pm$0.10, 2.18$\pm$0.21 and 0.17 dex.

Overplotted in \Fig{fig:btf} are the best fit BTFRs from the most up-to-date observations \citep[][Lelli et al. 2015, submitted to ApJL]{mcgaugh12BAR} \lcdm-based theoretical models \citep{duttonBTF,tg11} and recent hydrodynamical simulations \citep{santos15}. 
The slope, intercept, and scatter of the BTFR represent one of the major predictions of MOND \citep{milgrom83}, whereas it has been argued that they are difficult to reproduce within a \lcdm\ context without fine-tuning \citep{mcgaugh05,mcgaugh12BAR}. We show here that a semi-empirical model, coupling observed scaling relations with \lcdm\ haloes, can generate a BTFR in good agreement with the observed one. The scatter of 0.17 dex is somewhat high compared to the observational intrinsic scatter ($\sim$0.11 dex, Lelli et al. 2015, submitted to ApJL). Whether this is a challenge for \lcdm, or is it rather pointing out the sensitivity of the BTFR to the measurements radius, is an open issue which deserves further investigation.

\begin{figure}
\vspace{-.1in}\hspace{-.2cm} \includegraphics[width=3.4in,height=2.8in]{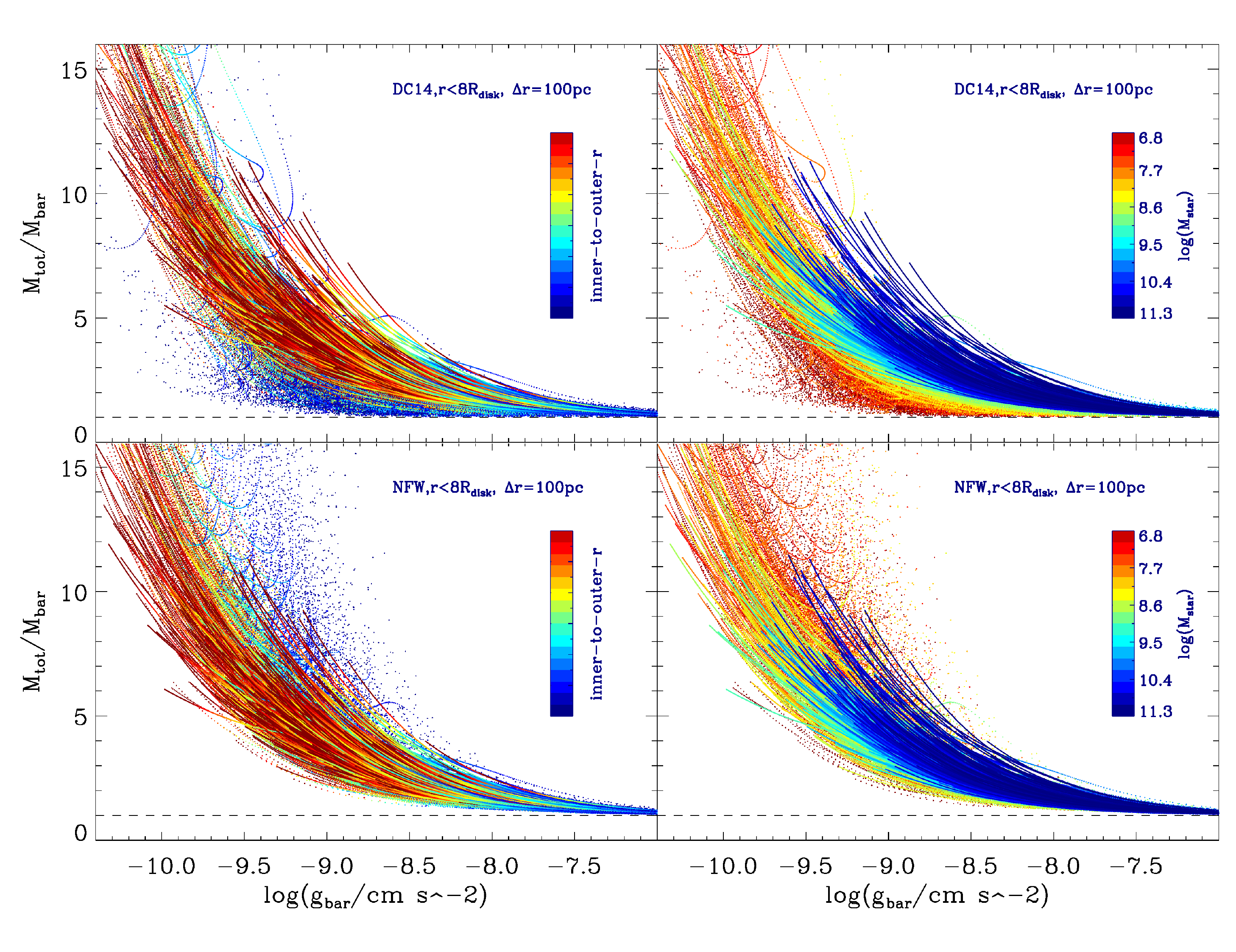}

  \caption{The MDAR of model galaxies in the mass range 6.8$\leqslant \rm log_{10}(M^{\star}/$\msun)$\leqslant$11.3 for DC14 (top) and NFW (bottom) profiles, measured out to a maximum radius of $\rm r=8R_{disk}$. The MDAR of each galaxy is color coded by radius (left) and stellar mass (right). The dashed black line shows the maximum baryonic contribution.}
\label{fig:MDAR_radius_halo} 
\end{figure}

The mass discrepancy, M$\rm_{tot}(r)$/M$\rm_{bar}(r)$, and the corresponding baryonic acceleration, $\rm g_{bar}(r)$, are computed for each model galaxy within the range $\rm r$$<$$\rm4R_{gas}$$=$$\rm8R_{disk}$, which is a conservative proxy for the outer-most radius measured in HI observations. In \Fig{fig:MDAR_radius_halo} we plot such mass discrepancy for the full sample of $\sim$1500 model galaxies against g$_{\rm bar}$, using uniformly spaced bins of 100 pc (different binning criteria produce less, or more, dense plots, without changing the general pattern shown in \Fig{fig:MDAR_radius_halo}).
In the top two panels the MDAR is shown for a DC14 profile, color coded according to the radius at which it is measured (left panel), where blue indicates the inner-most radii and red the outer-most ones, and according to the stellar mass (right panel), from dwarf galaxies in red to massive spirals in blue. The same color coding scheme is used for a NFW model (bottom panels). Each galaxy is represented by a series of points. Low mass galaxies overlaps with high mass ones to produce the MDAR relation. 

\Fig{fig:MDAR_radius_halo} illustrates that there is a clear difference between DC14 and NFW models.
Low mass galaxies with 7$\lesssim$$\rm log_{10}$($\rm M^{\star}$/\msun)$\lesssim$9 have a shallow, core-like inner profile ($\gamma$$\lesssim$0.5) in the DC14 model, resulting in a lower mass discrepancy in their inner region than the corresponding cuspy NFW case.
As a consequence the shape of the MDAR in the DC14 model between  -10$\lesssim$$\rm log_{10}(g_{bar}/cms^{-2}$)$\lesssim$-8.5 is different than the NFW one. On the contrary, NFW low mass galaxies have a central cusp and high values of M$\rm_{tot}(r)$/M$\rm_{bar}(r)$ due to the increased contribution of DM to their center.

\begin{figure}
  \hspace{-.3cm}\includegraphics[width=3.5in]{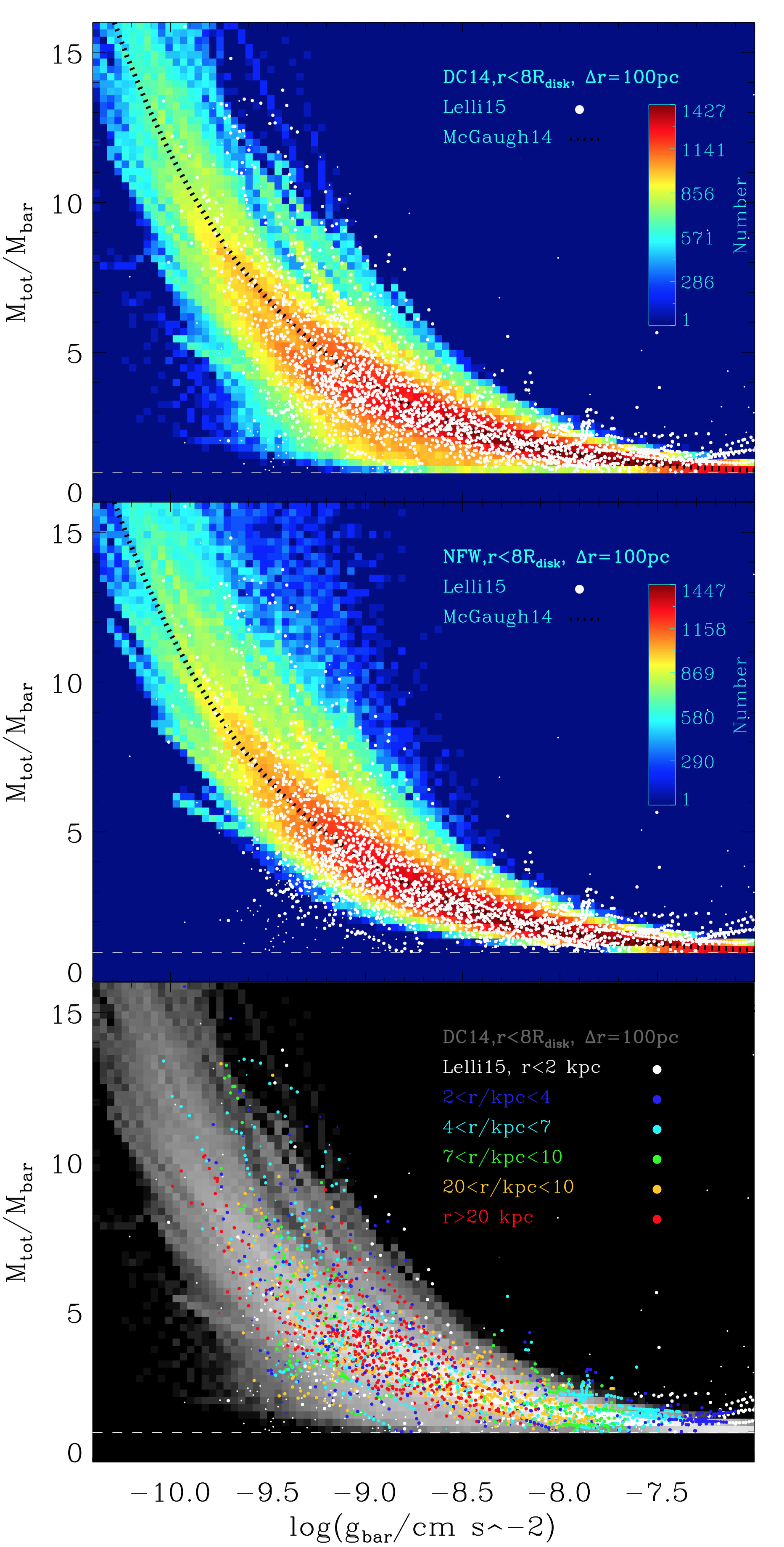}
      \caption{Density map of the MDAR generated by model galaxies that follow the same stellar mass distribution of the observational sample. Results for the DC14 profile (top and bottom panels) and the NFW one (central panel) are shown, out to a radius  $\rm r$=$\rm8R_{disk}$ and using bins of 100 pc. Filled circles represent the observed MDAR derived from the SPARC sample of 146 galaxies. In the bottom panel the DC14 heat plot is shown in black and white, with observational points color-coded by their measured radius.}
      \label{fig:MDAR_1000} 
\end{figure}

 A direct comparison with observations is offered in \Fig{fig:MDAR_1000}, whose three panels show the density map of the MDAR generated by our model galaxies using a DC14 profile (top and bottom panel) and a NFW one (central panel). Once again we used uniformly spaced bins of 100 pc, although different bin sizes do not alter our results.
Overplotted as filled circles are the observational data referring to the 146 galaxies from the SPARC dataset, considering only points with a relative error on the measured rotation curve errV/V$\leqslant$0.5.
Big circles indicate the most accurate data, with errV/V$\leqslant$0.1, while small circles represent  data whose relative error is 0.1$\leqslant$errV/V$\leqslant$0.5. The dashed line is the observational fit  from \citet{mcgaugh14}.
 
Our semi-empirical-\lcdm-based model is able to fully account for both the shape and the observed scatter in the relation. Comparing the first two panels, it is also clear that the DC14 profile encompasses the full range of  the observed MDAR, while the NFW one fails to account for several points in the low acceleration-low mass discrepancy space. This is the region where dwarf galaxies with inner cores are found in the DC14 model (see \Fig{fig:MDAR_radius_halo}). 
To further investigate this issue we show, in the bottom panel of  \Fig{fig:MDAR_1000}, the same DC14 heat-map with observational points  color-coded according to the radius at which they have been measured. 
The observational points falling in the bottom left area of the MDAR plot come from dwarf galaxies with \mstar$<$$10^{9.5}$\msun\ and mostly from their inner region, in agreement with DC14 expectations. We predict that dwarf galaxies falling in this area of the MDAR should have a central core-like distribution, which a universal NFW halo model cannot reproduce.

\section{Conclusions}
The MDAR, i.e. the relation between total-to-baryonic mass ratio as a function of baryonic acceleration at different radii of a galaxy, has been historically seen as a challenge to the \lcdm\ paradigm \citep{mcgaugh14,wu2015}. Such tight relation and its  scatter can be instead easily reproduced within a MOND context, in which a characteristic acceleration scale exists \citep{milgrom83}.

In this \textit{Letter} we have shown that the MDAR can be reproduced using a semi-empirical model in a \lcdm\ scenario, using theoretically-motivated DM haloes  populated with realistic disc galaxies that follow observed structural relations.
We created  model  galaxies whose distribution in stellar, gas, baryonic mass, disk scale length and central surface brightness  closely match the observational SPARC dataset (Lelli et al. in prep.), enabling accurate comparisons between theory and observation.

The MDAR generated by galaxies embedded in DM  haloes that follow the mass-dependent DC14 profile \citep{dicintio2014a,dicintio2014b} is able to account for the overall shape and scatter of the observed MDAR, from dwarfs to massive spirals. Instead, employing the universal NFW \citep{navarro96} profile fails to reproduce the MDAR generated by dwarf galaxies with \mstar$\lesssim$$10^{9.5}$\msun, a further evidence pointing toward the existence of cores in such objects. More accurate data of the inner kpc of dwarf galaxies could better discriminate the two models.
Regardless, both models indicate that the MDAR can be reproduced in a \lcdm\ context once the observed structural relations of galaxies are taken into account.

Our model can also reproduce the slope and normalization of the BTFR within the observational errors \citep{mcgaugh12BAR}, in line with some \lcdm\ semi-analytic models \citep{duttonBTF,tg11} and state-of-the-art hydrodynamical simulations \citep{santos15}. This implies that the global baryonic fractions from abundance-matching techniques \citep[e.g.][]{kravtsov14} are roughly consistent with those implied by the BTFR. As in previous models, however, the scatter in the BTFR is slightly higher than the observed intrinsic scatter, an issue that deserves further investigations.

 This work extends the results of previous MDAR studies based on the analysis of hydrodynamical simulations \citep{santos15} and  semi-analytic models \citep{vdb00}, largely confirming that once the general patterns described by galactic structural relations are properly taken into account, capturing the physics of galaxy formation, the MDAR and the BTFR arise in a \lcdm\ context.
\section*{Acknowledgements}
ADC is supported by the DARK - Carlsberg independent fellowship program. ADC thanks N. Amorisco and C. Brook for fruitful discussions. The work of FL is made possible through the support of a grant from the John Templeton Foundation.
\bibliographystyle{mn2e}
\bibliography{archive}


\label{lastpage}

\end{document}